\documentclass{article} 
\usepackage{iclr2021_conference,times}

\usepackage{amsmath,amsfonts,bm}

\def\eqref#1{equation~\ref{#1}}

\def\1{\bm{1}}

\DeclareMathAlphabet{\mathsfit}{\encodingdefault}{\sfdefault}{m}{sl}
\SetMathAlphabet{\mathsfit}{bold}{\encodingdefault}{\sfdefault}{bx}{n}

\usepackage{hyperref}
\usepackage{url}
\usepackage{fontawesome5}
\usepackage{cleveref}
\usepackage{times}
\usepackage{latexsym}
\usepackage{arydshln} 
\usepackage{amsmath}
\usepackage{amsfonts}
\usepackage{bm}
\usepackage{graphicx}
\usepackage{multirow}
\usepackage{makecell}
\usepackage{array}
\usepackage[para]{threeparttable}
\usepackage{booktabs}
\usepackage{tabularx}
\usepackage{xspace}
\usepackage{xcolor}
\usepackage{caption}
\usepackage{subcaption}
\usepackage{longtable}
\usepackage{bbm}
\usepackage{enumitem}
\usepackage{algorithm}
\usepackage{algpseudocode}
\usepackage{geometry}
\usepackage{comment}
\usepackage{tablefootnote}
\usepackage{color}

\usepackage[colorinlistoftodos]{todonotes}

\usepackage[T1]{fontenc}

\usepackage[utf8]{inputenc}

\usepackage{microtype}

\usepackage{inconsolata}

\newcommand\newsubcap[1]{\phantomcaption%
       \caption*{\figurename~\thefigure(\thesubfigure): #1}}

\hypersetup{hidelinks}

\crefrangelabelformat{section}{#3#1#4--#5\crefstripprefix{#1}{#2}#6}
\makeatletter
\newcommand{\crefnames}[3]{%
  \@for\next:=#1\do{%
    \expandafter\crefname\expandafter{\next}{#2}{#3}%
  }%
}
\makeatother

\crefnames{part,chapter,section}{\S}{\S}

\crefformat{section}{§#2#1#3}

\title{Back to Basics: A Simple Recipe for Improving\\ Out-of-Domain Retrieval in Dense Encoders}

\author{%
  Hyunji Lee$^{\hspace{.1em}{\boldsymbol{\kappa}}}$\thanks{\hspace{1mm}Work performed during internship at AI2.} \quad
  Luca Soldaini$^{\hspace{.1em}\boldsymbol{\alpha}}$ \quad
  Arman Cohan$^{\hspace{.1em}\boldsymbol{\gamma},\boldsymbol{\alpha}}$ \quad
  Minjoon Seo$^{\hspace{.1em}\boldsymbol{\kappa}}$ \quad 
  Kyle Lo$^{\hspace{.1em}\boldsymbol{\alpha}}$ \quad \\
  $^{\kappa\hspace{.1em}}$KAIST AI \quad
  $^{\alpha\hspace{.1em}}$Allen Institute for AI \quad
  $^{\gamma\hspace{.1em}}$Yale University\\
  {\tt  hyunji.amy.lee@kaist.ac.kr} \quad {\tt \{lucas, kylel\}@allenai.org}\\
}

\iclrfinalcopy 
\begin{document}

\maketitle

\begin{abstract}

Prevailing research practice today often relies on training dense retrievers on existing large datasets such as MSMARCO and then experimenting with ways to improve zero-shot generalization capabilities to unseen domains.
While prior work has tackled this challenge through resource-intensive steps such as data augmentation, architectural modifications, increasing model size, or even further base model pretraining, comparatively little investigation has examined whether the training procedures themselves can be improved to yield better generalization capabilities in the resulting models.
In this work, we recommend a simple recipe for training dense encoders: Train on MSMARCO with parameter-efficient methods, such as LoRA, and opt for using in-batch negatives unless given well-constructed hard negatives. We validate these recommendations using the BEIR benchmark and find results are persistent across choice of dense encoder and base model size and are complementary to other resource-intensive strategies for out-of-domain generalization such as architectural modifications or additional pretraining.
We hope that this thorough and impartial study around various training techniques, which augments other resource-intensive methods, offers practical insights for developing a dense retrieval model that effectively generalizes, even when trained on a single dataset.

\vspace{1em}
\centerline{{\raisebox{-2.5pt}{\Large\faGithub{}}} \hspace{.5em}\href{https://github.com/amy-hyunji/lora-for-retrieval}{\texttt{github.com/amy-hyunji/lora-for-retrieval}}}
\end{abstract}

\section{Introduction}

\label{sec:intro}

Dense neural retrieval methods have been proven to be generally effective in many Information Retrieval (IR) tasks~\citep{Karpukhin2020DensePR,Izacard2021TowardsUD, Ni2021SentenceT5SS}. 
These methods use learned neural encoders to obtain dense vector representations of text and the relevance of passages for any given query is estimated by computing the dot product between their encodings.
Dense approaches can outperform traditional retrieval techniques (\textit{e.g.}, BM25~\citep{Robertson1976bm25}), as they estimate similarity beyond syntactic matching~\citep{Lin2022BertBeyond}.

Neural retrieval models are effective rankers in domains for which large supervised datasets exist (\textit{e.g.}, MSMARCO~\citep{Campos2016MSMA} or Google NQ~\citep{Kwiatkowski2019NaturalQA}).
Conversely, they might struggle to generalize to settings they have not been trained on, leading to challenges in handling out-of-domain tasks~\citep{Thakur2021BEIRAH,Ren2022ATE,simon2023MsShift}.
In most real-world applications, supervision data is not available; 
whereas, retrieval models play a key role in the nascent field of augmented language models across many new exciting scenarios~\citep{Mialon2023AugmentedLM}. 
Thus, it is essential to analyze techniques that can improve generalization to unseen domains.

Many approaches have been proposed to tackle out-of-domain generalization.
For example, \textbf{data augmentation} approaches use weak supervision or auxiliary systems to bridge to unseen tasks~\citep{Dai2022PromptagatorFD, Bonifacio2022InParsDA, SaadFalcon2023UDAPDRUD, Lin2023HowTT}.
Other works introduce \textbf{novel architectures} that assess relevance at the token-level multi embeddings rather than employing a single embedding per passage~\citep{Khattab2020ColBERTEA, Formal2021SPLADEVS, Lee2023RethinkingTR}. 
Moreover, empirical observations suggest that \textbf{increasing the model size} leads to better out-of-domain performance~\citep{Ni2021LargeDE}.
While these approaches show significant improvements, they require \textit{additional} resource-intensive steps: data augmentation requires additional steps of generating new datasets, fine-grained token-level interaction requires higher inference costs with a large storage footprint, and larger model size requires more GPU memory during training and inference.
Finally, recently proposed approaches use contrastive losses to \textbf{pretrain} domain-specific encoders without explicit supervision~\citep{Izacard2021TowardsUD, Xu2022LaPraDoRUP}. 
These methods, while more effective than statistical IR techniques, still underperform supervised rankers unless they are then also fine-tuned on large supervised datasets like MSMARCO.
In fact, despite being out-of-distribution for many real-world tasks, large supervised collections remain critical to improving zero-shot retrieval, particularly for larger and well-trained rankers~\citep{Ni2021LargeDE,Rosa2022InDO,Lin2023HowTT,Weller2023WhenDG}.

Despite the vast body of work on improving out-of-domain generalization through resource-intensive steps like data augmentation, novel architectures, and pretraining, we notice comparatively less work has been done on the \textbf{training strategies} themselves commonly used to fine-tune rankers on large supervised datasets.
In this work, we aim to answer the following question: \textbf{when training dense models on large data collections, what procedures lead to better out-of-domain retrieval performance?}
In particular, we aim to address the following research questions: 
\begin{itemize}[noitemsep,topsep=0pt,leftmargin=*]
    \item \textbf{(RQ1)} Do parameter-efficient fine-tuning (PEFT) methods, such as LoRA~\citep{Hu2021LoRALA}, improve performance on out-of-domain tasks?
    \item \textbf{(RQ2)} How might we modify the design of our batches for better out-of-domain performance?
    \item \textbf{(RQ3)} To what extent do our recommendations complement other resource-intensive techniques that improve out-of-domain generalization?
\end{itemize}
\noindent by identifying key design decisions for training dense retrieval models and conducting a series of carefully designed experiments that isolate the effects of these various decisions (\cref{sec:experiments}).

Addressing RQ1 in \S\ref{sec: rq1}, we find that \textbf{LoRA}, one of the most widely-used PEFT techniques, leads to better out-of-domain generalization performance compared to \textbf{full parameter tuning}. Simultaneously, we validate an intuitive tradeoff---full parameter tuning still outperforms LoRA on in-domain settings. Nevertheless, we provide further analysis showing that even when considering this \textbf{tradeoff} between in- and out-of-domain performance, LoRA may provide more than it gives up. We recommend LoRA as a suitable training approach when training a model that expects high performance in both in-domain and out-of-domain settings. 

Further, addressing RQ2 in \S\ref{sec: rq2}, we find that contrary to their well-established benefit in in-domain settings, mined \textbf{hard negatives} may hurt out-of-domain retrieval performance unless selected with great care. 
On the other hand, increasing the number of \textbf{in-batch negatives} is consistently beneficial for out-of-domain performance, a finding that can be opportunistically employed by adopting PEFT as our fine-tuning strategy. Specifically, under identical GPU configurations, increasing the in-batch size typically yields more robust performance compared to adding hard negatives.

Finally, addressing RQ3 in \S\ref{sec: rq3}, we find that our learnings complement other popular yet resource-intensive techniques for enhancing out-of-domain performance, such as adopting \textbf{larger base models}, novel retriever architectures (e.g., \textbf{late interaction models}), and additional \textbf{contrastive pretraining} of the base model. 

Our results show consistent trends across several encoder-only base models and common dual-encoder retriever architectures.
Combining the findings from RQ1 and RQ2, we speculate that common full parameter fine-tuning practices are prone to \textbf{overfitting} large popular datasets like MSMARCO. 
Finally, taking all our findings together, our work provides simple, actionable takeaways that yield better out-of-domain generalization for neural retrieval models that we recommend as complements to other resource-intensive methods.

\section{Background and Related Work}
\paragraph{Out-of-domain generalization in information retrieval}
Many data augmentation techniques have been proposed as a means to offset limited training data availability~\citep{Dai2022PromptagatorFD, Bonifacio2022InParsDA, SaadFalcon2023UDAPDRUD, Lin2023HowTT}. 
Fully unsupervised techniques can also be used to circumvent the lack of domain-specific supervised data~\citep{Izacard2021TowardsUD, Xu2022LaPraDoRUP}. 
Finally, modifications to the model itself have been explored, such as combining sparse retrieval~\citep{Formal2021SPLADEVS, Gao2021COILRE}, late-interaction learning~\citep{Lee2023RethinkingTR, Khattab2020ColBERTEA}, or using larger encoder models~\citep{Ni2021LargeDE,Neelakantan2022TextAC,Ma2023FineTuningLF}.
However, as mentioned in \cref{sec:intro}, all methods come with computational trade-offs:
they might require additional (expensive) training steps or have slower inference speeds. 
Therefore, our study aims to investigate various approaches that maximize the advantages of the dense retrieval approach while improving its generalizability performance, addressing these practical challenges.

\paragraph{Parameter efficient fine-tuning in information retrieval}

While standard fine-tuning of neural models typically entails training all model parameters, recent studies highlight the advantages of parameter-efficient fine-tuning (PEFT)~\citep{Houlsby2019ParameterEfficientTL, Hu2021LoRALA, BenZaken2021BitFitSP}. 
PEFT selectively updates a subset of model parameters or adds additional ones while keeping existing parameters fixed. 
These approaches offer several benefits, including reduced storage requirements, shorter training times, and lower GPU memory costs. 
Moreover, by not updating or only partially updating original parameters, PEFT helps prevent catastrophic forgetting and maintains robust performance in continual training~\citep{Wang2020KAdapterIK, Jin2021LifelongPC, Yoon2023ContinuallyUG, Jang2021TowardsCK}.
Due to the benefits and its competitive performance compared to full parameter tuning across various tasks, PEFT is widely used in machine learning~\citep{Liu2022FewShotPF,Ustun2022WhenDP}.

In information retrieval, the standard of full parameter tuning~\citep{Karpukhin2020DensePR, Lee2022GenerativeMR, Izacard2021TowardsUD, Formal2021SPLADEVS, Xiong2020AnsweringCO} is also giving way as PEFT gains traction.
\citet{Litschko2022ParameterEfficientNR} examined the use of PEFT for multilingual information retrieval;
\citet{Ma2022ScatteredOC} studied the use of PEFT to improve in-domain search capabilities;
\citet{Pal2023ParameterEfficientSR} applied PEFT to sparse retrieval systems, while \citet{Jung2021SemiSiameseBN} investigated improving hybrid retrieval;
\citet{Yoon2023ContinuallyUG} showed that PEFT can help adapt generative retrieval systems to new corpora.
\citet{Tam2022ParameterEfficientPT} is closest to our work in that they also study PEFT for out-of-domain generalization. 
We distinguish our work from theirs in several ways: (1) the scope of our study extends beyond the application of PEFT as we also consider the role of batch design (e.g., in-batch and hard negatives, \S\ref{sec: rq2}) and the effect of base models (e.g., model size, continued pretraining, \S\ref{sec: rq3}), (2) our experimental methodology (\S\ref{sec:experiments}) controls for differing amounts of training data seen by the base model; that is, we train different dense retrievers within an experiment group from the same base model whereas \citet{Tam2022ParameterEfficientPT} use different public model checkpoints fine-tuned on different datasets for different retrievers,
and (3) for PEFT method, we focus on LoRA, which they did not include in their work; they instead focus on P-tuning v2~\citep{liu-etal-2022-p} for zero-shot out-of-domain retrieval evaluations, as well as perform a wider sweep of different PEFT methods for in-domain evaluation, which we did not perform.

\paragraph{Batch design in information retrieval} 
When training dense encoders in a contrastive manner, it is not feasible to compute the loss across a corpus at each training step;
instead, the loss is computed over a smaller subset of positive and negative pairs. 
Consequently, many works adopt sampling strategies aimed at improving how  
training batches are constructed~\citep{Zhong2022TrainingLM, Lee2019ContextualizedSR, Min2022NonparametricML, Lee2022NonparametricDF, Qu2020RocketQAAO}. 
When organizing each batch, two aspects are widely known to be key for retrieval performance: 
(1) using relevant passages for one query as contrastive samples for other queries in the same batch, known as \textbf{``in-batch'' negatives}~\citep{Lindgren2021EfficientTO, Xiong2020AnsweringCO,Gillick2019LearningDR}, 
and (2) mining additional passages that are challenging to distinguish from relevant passages, known as \textbf{``hard'' negatives}~\citep{Karpukhin2020DensePR, Izacard2021TowardsUD, Luan2020SparseDA, Qu2020RocketQAAO, Wu2019ScalableZE}.  
Although the role of batch design has been widely studied for in-domain scenarios, there has not been an exploration of how such strategies translate to out-of-domain performance. 
Many works focusing on out-of-domain generalization tend to report the use of hard negatives without associated investigation to validate their use~\citep{Lee2023RethinkingTR, Izacard2021TowardsUD, Khattab2020ColBERTEA, Gao2021COILRE}. We believe our work is one of the first to question this practice as the de facto standard.

\section{Experimental Methodology}
\label{sec:experiments}

We carefully design an experimental procedure to study our various hypotheses. 
Specifically, we identify several key decision points when building a dense retrieval model. We make sure to define decision points such that: (1) we believe different design options will meaningfully impact end retrieval performance, and (2) we have the ability to experiment with different design variations at a single decision point while keeping others fixed, thereby allowing us to study the \emph{isolated} effect of those decisions at that single decision point.
In this work, our selected decision points are: (1) pretrained base model, (2) dense retriever architecture, (3) fine-tuning strategy, (4) how batches are constructed, and (5) datasets for training and testing.\footnote{We recognize that many of these design options are tightly coupled and difficult to fully study in isolation of each other. For example, a particular choice of retriever architecture will preclude certain base model choices as well as certain fine-tuning strategies.} 

\paragraph{Pretrained base models} 
We center our experimentation around BERT~\citep{devlin-etal-2019-bert}, the most popular choice of encoder-only base model~\citep{Karpukhin2020DensePR, Izacard2021TowardsUD, Khattab2020ColBERTEA}.
Additionally, focusing on BERT gives us the ability to study how controlled changes in the base model affect retrieval performance. For example, one can repeat an experimental run using different variants of BERT weights: (1) Different model sizes (e.g., Tiny, Small, Base, Large) to study the effect of scale, (2) RoBERTa~\citep{Liu2019RoBERTaAR} to study variation as a result of a different pretraining strategy rather than significant model architectural changes, and (3) Contriever~\citep{Izacard2021TowardsUD} to study variation as a result of retrieval-motivated continued pretraining using a contrastive loss.

\paragraph{Dense retriever architectures}

There are various architectural designs of encoder-only retrievers worth studying.
In particular, we consider \textbf{\texttt{dual encoder}} architectures, which use an encoder-only model to embed queries and documents such that their pairwise relevance can be derived by proximity (e.g., cosine similarity) in the shared embedding space.
In our work, we focus on three widely used designs. 
In the \textbf{\texttt{asymmetric}} dual encoder, the weights of the query and document encoders are not shared. Following the architecture design of \citet{Karpukhin2020DensePR}, we use the first token embedding (the CLS token embedding) as the representative embedding.
In the \textbf{\texttt{symmetric}} dual encoder, the weights of the query and document encoders are shared. Following the architecture design of \citet{Izacard2021TowardsUD}, we use the mean embedding (average of all token embeddings) as the representative embedding.
In the \textbf{\texttt{late interaction}} dual encoder~\citep{Lee2023RethinkingTR, Khattab2020ColBERTEA}, we use \emph{multiple} token embeddings as representative embeddings, unlike the symmetric and asymmetric dual encoders which use a \emph{single} representative embedding. We follow the architecture design of \citet{Khattab2020ColBERTEA} closely, including sharing the weights of the query and document encoders and use of the MaxSim operation to score similarity between each query against a bag of documents. 

\paragraph{Fine-tuning strategies}
We consider both full parameter tuning (FT) and PEFT for fine-tuning experiments. 
Among various PEFT methods~\citep{BenZaken2021BitFitSP, Liu2022FewShotPF, Ma2023FineTuningLF}, we focus on the low-rank adaptation (LoRA)~\citep{Hu2021LoRALA} method due to its wide usage~\citep{Dettmers2023QLoRAEF, Chen2023LongLoRAEF, Xu2023QALoRAQL}. 
LoRA keeps the pretrained model parameters fixed and integrates trainable rank decomposition matrices into each layer of the Transformer architecture. 
A key advantage of LoRA over other PEFT methods is that it does not increase inference latency, as it combines the trained parameters with the original weights during inference. We chose a rank of 7 and an alpha of 32, approximately 0.25\% of the original parameter count as trainable parameters.

\paragraph{Batch designs}

We first consider our options for handling \textbf{\texttt{mined hard negatives}}. 
While there is a line of research that shows adding hard negatives mined through a heavy distillation process improves performance~\citep{Santhanam2021ColBERTv2EA, Formal2021SPLADEVS,Ren2021RocketQAv2AJ}, this is resource-intensive and not broadly accessible. In this work, we focus on simple yet widely-used techniques: (1) \textbf{\texttt{BM25}}, (2) \textbf{\texttt{model self-distillation}} during training~\citep{Karpukhin2020DensePR, Izacard2021TowardsUD, Khattab2020ColBERTEA}, which have consistently shown to improve in-domain performance and does not have high dependency on other dual encoder models, (3) a combination of the two, (4) using \textbf{\texttt{dataset-provided}} hard negatives,\footnote{For NaturalQuestions, we use the version of the dataset released by \citet{Karpukhin2020DensePR} to utilize their hard negatives. For MSMARCO, we use the version from its official website and similarly use its provided hard negatives.} and (5) of course, the option to use \textbf{\texttt{no hard negatives}}.

Regarding \textbf{\texttt{in-batch negatives}}, this is driven by adjusting per-GPU batch size where each example in the batch is a positive query. PEFT methods take up less GPU memory to hold the model, thereby freeing up more space for larger per-GPU batch sizes. While per-GPU memory limitations can be overcome using gradient accumulation in many settings when training retrieval models, the use of in-batch negatives is practically limited to per-GPU batch size and not easily overcome through techniques like gradient accumulation without significant engineering and computation overhead~\citep{gao-etal-2021-scaling}. As such, we define a fixed per-GPU batch size \textbf{\texttt{(B)}} (see hyperparameters below), as well as settings for twice \textbf{\texttt{(2B)}} and four times \textbf{(\texttt{4B})} larger per-GPU batch sizes.

\paragraph{Datasets}
Focusing on evaluating whether models trained on large supervised datasets can generalize to out-of-domain tasks, for \textbf{\texttt{training}}, we use NaturalQuestions~\citep{Kwiatkowski2019NaturalQA} and MSMARCO~\citep{Campos2016MSMA}, two popular large datasets that have been used successfully used for this purpose~\citep{Thakur2021BEIRAH, Lee2023RethinkingTR, Izacard2021TowardsUD, Gao2021COILRE}.

For \textbf{\texttt{testing}}, we evaluate over 14 different datasets from the BEIR benchmark, which have been used in works studying out-of-domain generalization of retrieval models~\citep{Khattab2020ColBERTEA, Ni2021LargeDE, Weller2023WhenDG, Tam2022ParameterEfficientPT}. We evaluate over TREC-COVID (TR)~\citep{Roberts2020TRECCOVIDRA}, NFCorpus (NF)~\citep{Boteva2016AFL}, NaturalQuestions (NQ)~\citep{Kwiatkowski2019NaturalQA}, HotpotQA (HO)~\citep{Yang2018HotpotQAAD}, FIQA-2018 (FI)~\citep{Maia2018WWW18OC}, ArguAna (AR)~\citep{Wachsmuth2018RetrievalOT}, Touche-2020 (TO)~\citep{Bondarenko2020OverviewOT}, Quora (QU), DBpedia (DB)~\citep{Hasibi2017DBpediaEntityVA}, MSMARCO (MS)~\citep{Campos2016MSMA}, SCIDOCS (SD)~\citep{Cohan2020SPECTERDR}, FEVER (FE)~\citep{Thorne2018FEVERAL}, Climate-FEVER (CL)~\citep{Diggelmann2020CLIMATEFEVERAD}, and SciFact (SF)~\citep{Wadden2020FactOF}.
Of course, when training on MSMARCO, we treat MSMARCO evaluation as \textbf{\texttt{in-domain}} and all others as \textbf{\texttt{out-of-domain}}; similarly when training on NQ.

\paragraph{Hyperparameters}
Following \citet{Karpukhin2020DensePR}, we trained with effective batch sizes of 128 for 40 epochs with 10\% as warmup steps for both asymmetric and symmetric dual encoders unless otherwise specified\footnote{In the case of symmetric dual encoder, due to resource limitations, we conducted experiments with the same hyperparameters as \citet{Karpukhin2020DensePR} rather than \citet{Izacard2021TowardsUD}, as \citet{Izacard2021TowardsUD} is trained using a much larger batch size (1024) and a longer training duration (approximately 77 epochs). Also, please note that replicating the official contriever with MSMARCO in \citet{Izacard2021TowardsUD} is challenging because the optimizer code and negatives used during the training step are not released.}. All experiments are conducted using 8 or fewer A6000 GPUs (40GB memory), making the per-GPU batch size of 16. 
We use checkpoints for all pretrained models from Huggingface~\citep{Wolf2019HuggingFacesTS}. 
Since the number of training parameters differs for LoRA and FT, we perform early hyperparameter search over different learning rates $\in \{$1e-4, 2e-4, 1e-5, 2e-5, 5e-5$\}$ for both the FT and LoRA settings, following various configurations from previous works~\citep{Maillard2021MultiTaskRF, Karpukhin2020DensePR, Izacard2021TowardsUD, Xiong2020AnsweringCO, Ni2021LargeDE}. 
We found that the optimal learning rate for LoRA tends to be higher than FT---2e-5 for FT and 2e-4 for LoRA. For late interaction dual encoder, we follow the configuration from \citet{Khattab2020ColBERTEA}.

\paragraph{Evaluation Metrics}
Following the widely used metrics in the BEIR benchmark, we report results in nDCG@10 which calculates the ranking of the top 10 retrieved documents. All results are calculated with the official BEIR evaluation code~\citep{thakur2021beir}.

\section{How should we train? Comparing LoRA with full fine-tuning}
\label{sec: rq1}

\begin{table*}[ht!]
\centering
\fontsize{6.5}{10}\selectfont
    \begin{tabular}{cc|ccccccccccccc|c|c}
    \toprule
    & & \multicolumn{14}{c}{\textbf{Out-Of-Domain (OOD)}}  &  \multicolumn{1}{c}{\textbf{In-Domain}}\\
    \midrule
    & & FI & TR & NF & QU & SD & SF & TO & AR & HO & DB & FE & CL & NQ & \emph{Avg} & MS \\
    \midrule
    \multicolumn{14}{l}{\textbf{Asymmetric Dual Encoder}} \\
    \midrule
    \multirow{2}{*}{w/o Neg}
    & FT &  18.6 & 64.2 & 25.7 & 72.4 & 7.2 & 45.3 & 26.8 & 31.1 & 44.1 & \textbf{44.1} & 57.5 & 13.8 & \textbf{35.7} & 37.4 & \textbf{31.1}    \\
    & LoRA &  \textbf{19.5} & \textbf{67.9} & \textbf{27.1} & \textbf{73.4} & \textbf{8.0} & \textbf{49.4} & \textbf{30.0} & \textbf{33.8}  & \textbf{45.4} & 43.5 & \textbf{58.7} & \textbf{15.8} & 34.2 & \textbf{39.0} & 30.2    \\
    \midrule
    \multirow{2}{*}{w/ Neg}
    & FT & 17.8 & 63.0 & 23.5 & 69.7 & 6.8 & 40.6 & 20.6 & 26.0 & 34.8 & 32.8 & 54.1 & 12.6 & \textbf{36.9}  & 33.8 & \textbf{33.2}  \\
    & LoRA & \textbf{19.5} & \textbf{66.4} & \textbf{27.1} & \textbf{73.0} & \textbf{7.8} & \textbf{48.6} & \textbf{29.2} & \textbf{33.5} & \textbf{44.7} & \textbf{42.2} & \textbf{57.8} & \textbf{15.2} & 34.1 & \textbf{38.4}  & 30.9   \\
    \midrule
    \multicolumn{15}{l}{\textbf{Symmetric Dual Encoder}} \\
    \midrule
    \multirow{2}{*}{w/o Neg}
    & FT &  22.9 & 36.4 & 27.7 & 83.4 & \textbf{13.2} & 54.6 & \text{13.5} & \textbf{41.0}  & \textbf{49.4} & 30.3 & 60.7 & \textbf{18.7}& 33.7  & 37.3 & \textbf{31.3}   \\
    & LoRA &  \textbf{30.1} & \textbf{39.0} & \textbf{33.8} & \textbf{88.2} & 11.5 & \textbf{65.0} & \textbf{21.7} & 34.7  & 41.8 & \textbf{30.9} & \textbf{70.9} & 18.2& \textbf{34.8}  & \textbf{40.0} & 29.4  \\
    \midrule
    \multirow{2}{*}{w/ Neg}
    & FT & 19.0 & \textbf{34.7} & 24.5 & \textbf{67.1} & 9.6 & 44.4 & 11.9 & 32.4 & \textbf{38.4} & 20.2 & 56.6 & 10.3 & \textbf{35.4} & 31.1 & \textbf{36.9}   \\
    & LoRA & \textbf{25.5} & 34.3 & \textbf{27.9} & 62.8 & \textbf{10.0} & \textbf{55.7} & \textbf{16.5} & \textbf{33.8} & 37.8 & \textbf{22.7} & \textbf{64.9} & \textbf{13.8} & 30.2 & \textbf{33.5}  & 34.6   \\
    \bottomrule
    \end{tabular}
\caption
     {\fontsize{7.5}{10}\footnotesize Overall performance of dual encoder models trained on MSMARCO on BEIR benchmark tasks. The highest scores between the pair of full fine-tuning (FT) and LoRA experiments are in \textbf{bold}. 
     For all encoder architectures, we see two trade-offs between in-domain and out-of-domain tasks (\textit{``Avg''} column): (1) FT exhibits higher in-domain performance but lower out-of-domain performance, and (2) incorporating hard negative (\textit{``w/ Neg''} rows) consistently boosts in-domain performance but reduces out-of-domain performance.}
\label{table: LoRA_vs_full}
\end{table*}

In this section, we examine the impact of training techniques on both out-of-domain and in-domain performance; 
namely, we compare LoRA, a parameter-efficient training method, to the traditional approach of fully fine-tuning all model parameters (FT).
We conduct our experiments on two dense retrieval architectures: asymmetric and symmetric dual encoders.

\paragraph{LoRA consistently shows higher performance in out-of-domain over FT}

Table~\ref{table: LoRA_vs_full}, compares
the performance of dual encoder models trained using FT and LoRA techniques on the BEIR benchmarks. 
Results show that, on average, retrieval models trained with LoRA outperform FT by 1.6 to 4.6 absolute points (4.3\% to 13.7\% relative) on out-of-domain datasets. 
This result is consistent across different architecture, and suggest that LoRA is a more effective training technique for maximizing out-of-domain performance; 
conversely, models fully fine-tuned exhibit better in-domain performance. 
We speculate that this difference is due FT models overfitting to in-domain distribution, thus making them less versatile across out-of-domain datasets.

\paragraph{
LoRA offers a better trade-off between in-domain and out-of-domain performance
}

\begin{table*}[ht!]
\centering
\fontsize{6.5}{10}\selectfont
    \begin{tabular}{cc|cc|cc}
    \toprule
    & & \multicolumn{2}{c}{\textbf{Out-Of-Domain (OOD)}}  &  \multicolumn{2}{c}{\textbf{In-Domain}}\\
    \midrule
    & & \emph{Avg} &\textsc{Rel. Diff (\%)}& MS&\textsc{Rel. Diff (\%)} \\
    \midrule
    \multicolumn{4}{l}{\textbf{Asymmetric Dual Encoder}} \\
    \midrule
    \multirow{2}{*}{w/o Neg} 
    & FT  & 37.4 & & 31.1 \\
    & LoRA & \text{39.0} & \textbf{\multirow{-2}{*}{+4.1\%}} & 30.2 & \multirow{-2}{*}{-3.0\%} \\
    \midrule
    \multirow{2}{*}{w/ Neg}
    & FT   & 33.8 && \text{33.2} & \\
    & LoRA & \text{38.4} & \textbf{\multirow{-2}{*}{+12.0\%}}  & 30.9 &\multirow{-2}{*}{-7.4\%}  \\
    \midrule
    \multicolumn{4}{l}{\textbf{Symmetric Dual Encoder}} \\
    \midrule
    \multirow{2}{*}{w/o Neg}
    & FT & 37.3 & &\text{31.3}  & \\
    & LoRA  & \text{40.0} & \textbf{\multirow{-2}{*}{+6.8\%}} & 29.4 & \multirow{-2}{*}{-6.5\%}  \\
    \midrule
    \multirow{2}{*}{w/ Neg}
    & FT & 31.1 & &\text{36.9} &  \\
    & LoRA & \text{33.5} & \textbf{\multirow{-2}{*}{+7.2\%}}  & 34.6  &\multirow{-2}{*}{-6.6\%}\\
    \bottomrule
    \end{tabular}
\caption
     {\fontsize{7.5}{10}\footnotesize \textsc{Rel. Diff (\%)} is the percentage change between LoRA and FT trained models on in-domain and out-of-domain datasets.
     Results show that the out-of-domain performance increase of LoRA over FT always more than offset the decrease in in-domain performance.
     This suggests that LoRA is a more effective approach for training models that consistently perform well in both in-domain and out-of-domain scenarios.}
\label{table: in_out}
\end{table*}

Table~\ref{table: in_out} quantifies the trade-off between better in-domain performance afforded by FT techniques against better out-of-domain generalization with LoRA.
It is evident that in all scenarios, the decrease in average performance for out-of-domain datasets is more pronounced than the gains in in-domain datasets. These findings indicate that LoRA (PEFT) is a more suitable approach for training models that performs well in both in-domain and out-of-domain settings.

\paragraph{Asymmetric and Symmetric Encoders achieve similar performance} 
When comparing the two dual encoder models, we note that the two architectures perform similarly regardless of the training technique used.
While on individual datasets one might significantly outperform the other, their average performance on out-of-domain tasks is within one absolute point. 
Therefore, due to their popularity~\citep{Ni2021LargeDE, Ni2021SentenceT5SS, Lin2023HowTT, Karpukhin2020DensePR}, we choose to use asymmetric dual encoders for the remainder of our work, unless otherwise noted.

\section{How should we train? Designing optimal batches for fine-tuning}
\label{sec: rq2}

In this section, we focus on the influence of training batch design on both in-domain and out-of-domain performance. This includes examining the effects of incorporating hard negatives, a technique commonly acknowledged for enhancing in-domain performance, as well as the impact of selecting different batch sizes.

\begin{table*}[ht!]
\centering
\fontsize{6.5}{10}\selectfont
    \begin{tabular}{cc|cc|cc}
    \toprule
    & & \multicolumn{2}{c}{\textbf{Out-Of-Domain (OOD)}}  &  \multicolumn{2}{c}{\textbf{In-Domain}}\\
    \midrule
    & &  \emph{Avg} &\textsc{Rel. Diff (\%)}& MS &\textsc{Rel. Diff (\%)}\\
    \midrule
    \multicolumn{5}{l}{\textbf{Asymmetric Dual Encoder}} \\
    \midrule
    \multirow{2}{*}{FT}
    & w/o Neg & 37.4& & \text{31.1} &  \\
    & w/Neg & 33.8 &\textbf{\multirow{-2}{*}{-10.7\%}} & \text{33.2} & \textbf{\multirow{-2}{*}{+6.3\%}}  \\
    \midrule
    \multirow{2}{*}{LoRA}
    & w/o Neg & \text{39.0} &  & 30.2 &    \\
    & w/ Neg & \text{38.4}  & \text{\multirow{-2}{*}{-1.6\%}}  & 30.9  & \text{\multirow{-2}{*}{+2.3\%}}   \\
    \midrule
    \multicolumn{5}{l}{\textbf{Symmetric Dual Encoder}} \\
    \midrule
    \multirow{2}{*}{FT}
    & w/o Neg &  37.3 &  & \text{31.3}  &   \\
    & w/ Neg& 31.1  & \textbf{\multirow{-2}{*}{-19.9\%}} & \text{36.9}  & \textbf{\multirow{-2}{*}{+15.7}}  \\
    \midrule
    \multirow{2}{*}{LoRA}
    & w/o Neg  & \text{40.0} &  & 29.4  &  \\
    & w/ Neg & \text{33.5}  &\text{\multirow{-2}{*}{-19.4\%}}   & 34.6   &  \text{\multirow{-2}{*}{+12.9\%}} \\
    \bottomrule
    \end{tabular}
\caption
     {\fontsize{7.5}{10}\footnotesize \textsc{Rel. Diff (\%)} represents the percentage change in performance due to using mined negatives.
     FT shows a more significant reduction in out-of-domain (OOD) performance and a higher increase in in-domain (IN) performance relative to LoRA.}
\label{table: neg.reduction}
\end{table*}

\paragraph{Mined hard negatives degrade out-of-domain performance for FT and LoRA models.}
Table~\ref{table: neg.reduction} shows that hard negatives, despite generally enhancing in-domain performance, consistently degrade out-of-domain nDCG scores.
We hypothesize this is due to the fact that models tend to over-adapt to the training dataset when adding hard negatives, making it challenging to generalize to datasets with differing distributions. 
In the experiments, we use negatives from the official MSMARCO dataset following \citet{Khattab2020ColBERTEA}\footnote{As some queries are missing negatives, we fill the queries with BM25 negatives.}.

We note that FT models are more influenced by hard negatives than models trained with LoRA. 
while they benefit more from the inclusion of mined negatives on in-domain tasks, 
their performance decreases more severely on out-of domain tasks.
The finding suggests that since FT trains a larger number of parameters than LoRA, it becomes more attuned to the given datasets, and further gets more affected by hard negatives.
This results in more pronounced improvements in in-domain performance, but at the cost of a larger decrease in out-of-domain performance.

\paragraph{Experimental evidence suggests hard negatives encourage overfitting to training data distribution.}

\begin{table*}[t!]
\centering
\fontsize{6.5}{10}\selectfont
    \begin{tabular}{c|cc|cc}
    \toprule
    & \multicolumn{2}{c}{\textbf{Asymmetric Dual Encoder}} & \multicolumn{2}{c}{\textbf{Symmetric Dual Encoder}} \\    
    \midrule
    &  FT & LoRA & FT & LoRA \\
    \midrule
    Similar Distribution (\%) & -8.6\% & -1.4\% & -16.7\% & -11.8\% \\
    Different Distribution (\%) & \textbf{-11.1\%} & \textbf{-1.7\%}  & \textbf{-18.8\%} & \textbf{-18.8\%} \\
    \bottomrule
    \end{tabular}
\caption
     {\fontsize{7.5}{10}\footnotesize 
     Relative change in performance from using hard negatives.
     We partition out-of-domain datasets in BEIR by how similar they are to MSMARCO. 
     Similarity is assesses by sampling 50 documents from each dataset, and comparing their average Contriever embeddings.
     Datasets that are most dissimilar to MSMARCO have consistently higher relative decrease in performance than the group of most similar datasets.}
\label{table: neg.dist}
\end{table*}

We set out to investigate our hypothesis---performance decline when incorporating hard negatives is due to over-specialization to a specific training dataset---by empirically assessing whether dateset that are most \textit{dissimilar} from training data are more severely affected.
To assess similarity between datasets, we randomly select 50 instances from a corpus of each dataset and compute the inner product of embeddings from contriever~\citep{Izacard2021TowardsUD}. The dataset most frequently identified as top-ranked, excluding its own dataset, was considered the most similar. After identifying the most relevant dataset for each dataset, we then grouped them together.
Repeating this process five times, we categorize datasets grouped with MSMARCO more than three times as similar distribution.
The BEIR datasets most similar to MSMARCO are trec-covid, NFcorpus, scidocs, scifacts, and arguana (see Appendix~\ref{app: domain_similarity} for details).

We summarize our findings in Table~\ref{table: neg.dist}. 
Results show that datasets that are most similar to MSMARCO exhibit a smaller performance drop when hard negatives are added.
When comparing the average drop rates between these groups, we observed that those in similar domains showed a lesser reduction (average of 3\%) compared to those grouped in different distributions. This suggests that training with hard negatives tends to overfit the model to its training dataset, reducing its effectiveness on datasets with different distributions.

\paragraph{Unlike mined negatives, using larger batch size increases both in-domain and out-of-domain performance.}

\begin{table*}[t!]
\centering
\fontsize{6.5}{10}\selectfont
    \begin{tabular}{cc|ccccccccccccc|c|c}
    \toprule
        && \multicolumn{14}{c}{Out-Of-Domain (OOD)} & \multicolumn{1}{c}{In-Domain} \\
    \midrule
         & Batch Size & FI & TR & NF & QU & SD & SF & TO & AR & HO & DB & FE & CL & NQ & \emph{Avg} & MS  \\
    \midrule
     FT  &B &  18.6 & 64.2 & 25.7 & 72.4 & 7.2 & 45.3 & 26.8 & 31.1 & 44.1 & {44.1} & {57.5} & 13.8 & \textbf{35.7}  & 37.4 & \text{31.1}   \\
     \midrule
     \multirow{3}{*}{LoRA} 
     & B &  {19.5} & 67.9 & {27.1} & 73.4 & {8.0} & {49.4} & {30.0} & {33.8}  & {45.4} & 43.5 & {58.7} & {15.8} & 34.2 & {39.0} & 30.2  \\
    & $2 \times \text{B}$ &  20.7 & 71.1 & 27.5 & \textbf{76.1} & 8.0 & 51.5 & \textbf{30.8} & 33.9 & 46.3 & 44.2 & 58.7 & 14.2 & 34.1& 39.8 & 31.0  \\
    & $4\times\text{B}$ & \textbf{20.9} & \textbf{71.4} & \textbf{28.3} & 75.5 & \textbf{8.1} & \textbf{52.6} & 30.2 & \textbf{34.3} & \textbf{47.2} & \textbf{44.8} & \textbf{59.7} & \textbf{14.7} & \text{35.0}& \textbf{40.2} & \textbf{31.8}  \\
    \bottomrule
    \end{tabular}
\caption
     {\fontsize{6.5}{10}\footnotesize Increasing batch size (increasing the number of in-batch negatives) consistently helps both in-domain and out-of-domain performance. }
\label{table: MSMARCO.inbatch}
\end{table*}

In Table~\ref{table: MSMARCO.inbatch}, we observe that using larger batch sizes, which include a greater number of in-batch negatives, enhances performance both within and outside the domain. This observation suggests that, under certain GPU configurations, to boost performance across both domains, increasing batch size could be a more effective strategy than incorporating hard negatives.
We hypothesize that this is because hard negatives often lead the model to over-adapt to certain distributions. On the other hand, in-batch negatives, which are typically random negatives, do not exhibit such a tendency. 
Moreover, since LoRA demands less GPU memory, it enables the use of larger batch sizes under the same GPU constraints compared to FT. In our experiments, LoRA with a batch size quadruple that of FT consumes a similar amount of GPU memory.

\section{How complementary are our findings with other resource-intensive methods for out-of-domain retrieval?}
\label{sec: rq3}

In this section, we investigate (1) the impact of increasing model size, (2) the use of late-interaction retrieval architectures, and (3) whether our findings still add benefit on top of more powerful base models that have undergone additional pretraining.

\begin{figure}[t!]
    \centering
    \includegraphics[width=0.9\textwidth]{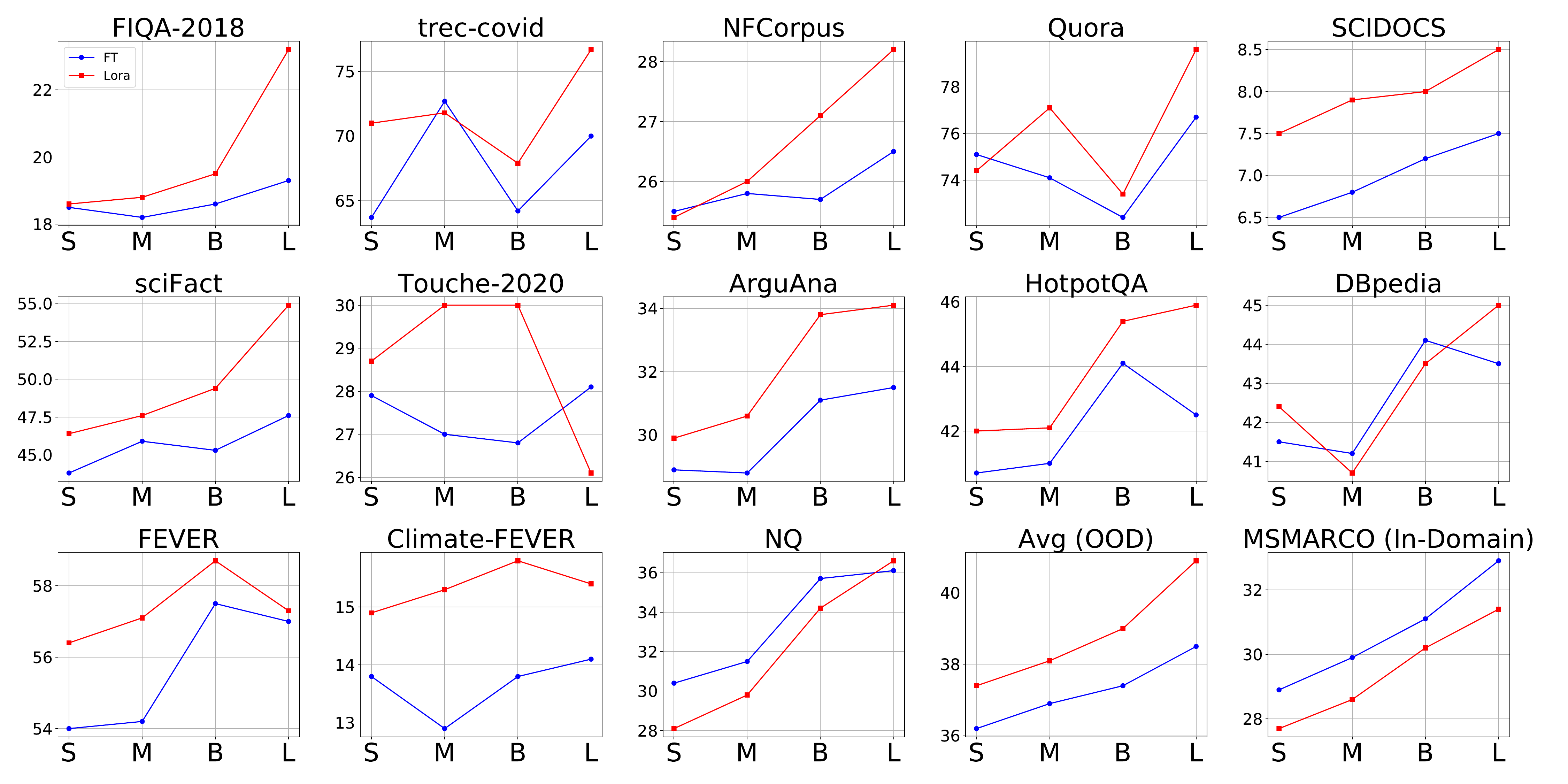}
    \caption{\fontsize{6.5}{10}\footnotesize Overall BEIR performance of different base model sizes of asymmetric dual encoder trained with MSMARCO (without hard negatives). Performance consistently increases with larger encoder models for both in-domain and out-of-domain. Notation in x-axis indicates \textbf{S}: Bert Small, \textbf{M}: Bert Medium, \textbf{B}: Bert Base, \textbf{L}: Bert Large}
    \label{fig: MSMARCO.basemodel}
\end{figure}

\begin{table*}[ht!]
\centering
\fontsize{6.5}{10}\selectfont
    \begin{tabular}{cc|ccccccccccccc|c|c}
    \toprule
    & & \multicolumn{14}{c}{\textbf{Out-Of-Domain (OOD)}}  &  \multicolumn{1}{c}{\textbf{In-Domain}}\\
    \midrule
   & & FI & TR & NF & QU & SD & SF & TO & AR & HO & DB & FE & CL & NQ & \emph{Avg} & MS \\    
    \midrule
    \multicolumn{14}{l}{\textbf{Bert Large}} \\
    \midrule
    \multirow{2}{*}{w/o Neg} 
    & FT &  19.3 & 70.0 & 26.5 & 76.7 & 7.5 & 47.6 & \textbf{28.1} & 31.5 & 42.5 & 43.5 & 57.0 & 14.1 & 36.1 & 38.5  & 32.9 \\
    & LoRA &  \textbf{23.2} & \textbf{76.7} & \textbf{28.2} & \textbf{79.6} & \textbf{8.5} & \textbf{54.9} & 26.1 & \textbf{34.1} & \textbf{45.9} & \textbf{45.0} & \textbf{57.3} & 15.4 & \textbf{36.6}& \textbf{40.9} & 31.4   \\
    \midrule
    \multirow{2}{*}{w/ Neg} 
    & FT &  17.5 & 69.5 & 23.3 & 74.6 & 7.2 & 44.4 & 22.7 & 30.7 & 40.5 & 42.8 & 54.5 & 11.7 & 37.3& 36.7 & \textbf{34.4}   \\
    & LoRA &  \textbf{22.7} & \textbf{75.0} & \textbf{28.2} & \textbf{78.5} & \textbf{8.4} & \textbf{54.3} & \textbf{25.2} & \textbf{31.9} & \textbf{42.3} & \textbf{44.3} & \textbf{56.5} & \textbf{15.5} & \textbf{34.6}& \textbf{39.8} & 32.1 \\
     \midrule
    \multicolumn{14}{l}{\textbf{RoBERTa Large}} \\
        \midrule
    \multirow{2}{*}{w/o Neg} 
    & FT &  24.5 & 65.9 & 27.9 & 77.8 & 9.0 & 48.1 & \textbf{30.2} & 32.1 & 41.5 & 40.6 & 58.7 & 14.7 & 34.5 & 38.9 & \text{34.0}  \\
    & LoRA & \textbf{29.0} & \textbf{73.7} & \textbf{28.8} & \textbf{78.6} & \textbf{9.4} & \textbf{55.2} & \text{26.2} & \textbf{37.8} & \textbf{43.7} & \textbf{42.7} & \textbf{59.9} & \textbf{18.4} & 35.5& \textbf{41.5} & 32.1  \\
    \midrule
    \multirow{2}{*}{w/ Neg} 
    & FT &  24.5 & 55.9 & 25.9 & \textbf{77.8} & 7.0& 48.1 & 22.2 & 32.1& 38.5 & 36.6 & 56.7 & 13.7 & 32.5 & 36.3& \textbf{35.9}  \\
    & LoRA & \textbf{28.0} & \textbf{70.2} & \textbf{28.7} & 76.1 & \textbf{8.3} & \textbf{53.2} & \textbf{25.9} & \textbf{36.7} & \textbf{42.5} & \textbf{41.9} & \textbf{58.9} & \textbf{16.8} & \textbf{36.8} & \textbf{40.3} & 34.5  \\
    \bottomrule
    \end{tabular}
\caption
     {\fontsize{6.5}{10}\footnotesize Overall BEIR performance of RoBERTa-large and BERT-large, two similar base model architectures and size but with different training strategies. (1) Performance tends to increase when using RoBERTa-large. (2) Our findings about the benefits of LoRA (\S\ref{sec: rq1}) and the possible detriment of hard negatives (\S\ref{sec: rq2}) hold here as well.}
\label{table: MSMARCO.RoBERTa}
\end{table*}

\paragraph{Impact of increasing model size: out-of-domain performance improves with better base model especially in LoRA}
We study whether our findings hold when increasing the size of the base model, such as in \citet{Ni2021LargeDE}. Figure~\ref{fig: MSMARCO.basemodel} shows results using BERT models in four different sizes (small, medium, base, large) finetuned on MSMARCO.
We can see that larger base model consistently leads to higher performance in all cases See Appendix Table~\ref{table: MSMARCO.basemodel} for numerical results. From this table, we also validate our earlier findings about the trade-offs between in-domain versus OOD performance when using LoRA versus FT and the possible detrimental effects of using hard negatives.

We further experiment by switching the base model from BERT-large to RoBERTa-large, a model trained with more optimized hyperparameters (Table~\ref{table: MSMARCO.RoBERTa}). We observe that when training with RoBERTa-large instead of BERT-large, both in-domain and out-of-domain performance show improvement. 
And of course, we re-validate our consistent findings about LoRA versus FT and hard negatives from earlier.

Looking deeper into the LoRA versus FT tradeoff, we can see that LoRA tends to benefit more by using a better base model (Table~\ref{table: model_size.dist}). 
This matches intuition as PEFT does not update many of the base model parameters; therefore, as the base model improves, so does the performance of a LoRA trained model as it can use much more information from the frozen parameters. FT, on the other hand, updates the whole base model and is more likely affected by forgetting and shifts in the distribution across all parameters.

\begin{table*}[ht!]
\centering
\fontsize{6.5}{10}\selectfont
    \begin{tabular}{c|ccc|ccc}
    \toprule
    & \multicolumn{3}{c}{OOD Avg} & \multicolumn{3}{c}{In-domain} \\    
    \midrule
    &  Medium & Base & Large & Medium & Base & Large \\
    \midrule
    FT & \textbf{1.9\%} & 3.3\% & 6.4\% & \textbf{3.5\%} & \text{7.6\%} & \text{11.6\%}  \\
    LoRA & \textbf{1.9\%} & \textbf{4.3\%} & \textbf{9.4\%} & 3.2\% & \textbf{9.0\%} & \textbf{13.4\%} \\
    \bottomrule
    \end{tabular}
\caption
     {\fontsize{7.5}{10}\footnotesize Performance using larger BERT model sizes relative to performance using BERT-small weights. (1) Performance improves monotonically with model size. (2) LoRA tends to benefit more from a larger base model compared to FT. 
     }
\label{table: model_size.dist}
\end{table*}

\paragraph{Use of token-level late-interaction models}

\begin{table*}[ht!]
\centering
\fontsize{6.5}{10}\selectfont
    \begin{tabular}{c|ccccccccccccc|c|c}
    \toprule
    & \multicolumn{14}{c}{\textbf{Out-Of-Domain (OOD)}}  &  \multicolumn{1}{c}{\textbf{In-Domain}}\\
    \midrule
    & FI & TR & NF & QU & SD & SF & TO & AR & HO & DB & FE & CL & NQ & \emph{Avg} & MS \\
    \midrule
     FT & 31.9 & \textbf{65.1} & 31.8 & \textbf{83.8} & 15.1 & 67.5 & 19.5 & 24.0 & 58.4 & \textbf{38.7} & 77.9 & 18.2 & \textbf{51.7} & 44.9 & \textbf{39.2}   \\
    LoRA & \textbf{32.5} & 64.4 & \textbf{32.7} & 83.3 & \textbf{15.6} & \textbf{68.5} & \textbf{21.6} & \textbf{37.4} & \textbf{61.1} & 33.6 & \textbf{78.6} & \textbf{19.1} & 51.2 & \textbf{46.1} & 37.2  \\  
    \bottomrule
    \end{tabular}
\caption
     {\fontsize{7.5}{10}\footnotesize Overall BEIR performance of a token-level late-interaction dual encoder following \citet{Khattab2020ColBERTEA} and trained on MSMARCO. Like all other models tested, we see a clear performance trade-off between FT, which is better for in-domain performance, and LoRA, which is better for out-of-domain performance.}
\label{table: colbert}
\end{table*}

We experiment with a late interaction dual encoder model following \citet{Khattab2020ColBERTEA}to see whether our findings persist in retrieval architectures that involve more resource-intensive steps.\footnote{Our FT numbers are obtained through a replication of \citet{Khattab2020ColBERTEA} experiments, which was necessary to make a fair comparison between LoRA and FT, as opposed to using their released weights. Nevertheless, our results are very close to the ones reported in \citet{thakur2021beir}, indicating our successful replication. For more details, see Appendix~\ref{app: colbert}.} 
Table~\ref{table: colbert} shows a similar trend to what we observed in asymmetric and symmetric dual encoders; LoRA surpasses FT in out-of-domain settings, whereas FT demonstrates superior performance in in-domain settings.

\paragraph{The effectiveness of employing models pre-trained on contrastive loss as the initial base model}

\begin{table*}[ht!]
\centering
\fontsize{6.5}{10}\selectfont
    \begin{tabular}{c|ccccccccccccc|c|c}
    \toprule
    & \multicolumn{14}{c}{\textbf{Out-Of-Domain (OOD)}}  &  \multicolumn{1}{c}{\textbf{In-Domain}}\\
    \midrule
    & FI & TR & NF & QU & SD & SF & TO & AR & HO & DB & FE & CL & NQ & \emph{Avg} & MS \\    
    \midrule
    \multicolumn{13}{l}{\textbf{Bert Base}} \\
    \midrule
    FT &  22.9 & 36.4 & 27.7 & 83.4 & \textbf{13.2} & 54.6 & \text{13.5} & \textbf{41.0}  & \textbf{49.4} & 30.3 & 60.7 & \textbf{18.7}& 33.7  & 37.3 & \textbf{31.3}  \\
    LoRA &  \textbf{30.1} & \textbf{39.0} & \textbf{33.8} & \textbf{88.2} & 11.5 & \textbf{65.0} & \textbf{21.7} & 34.7  & 41.8 & \textbf{30.9} & \textbf{70.9} & 18.2& \textbf{34.8} & \textbf{40.0} & 29.4  \\
    \midrule
     \multicolumn{14}{l}{\textbf{Bert Base with Contrastive Pretraining (Contriever)}} \\
    \midrule
    FT & \textbf{26.9} & 40.1 & 30.5 & 84.4 & 14.9 & 64.4 & 13.4 & \textbf{40.9} & 60.6 & 37.5 & 68.8 & \textbf{20.4} & \textbf{38.9} & 41.7   & \textbf{32.8}\\
    LoRA & \textbf{26.9} & \textbf{44.6} & \textbf{33.9} & \textbf{88.7} & \textbf{15.8} & \textbf{65.7} & \textbf{20.6} & 36.2 & \textbf{62.3} & \textbf{38.3} & \textbf{71.5} & 19.0 & 37.1 & \textbf{43.1} & 31.5  \\
    \bottomrule
    \end{tabular}
\caption
     {\fontsize{6.5}{10}\footnotesize  
     Comparison between two encoder models derived from a BERT-base and Contriver checkpoint.
     Contriver was obtained by further training a BERT-base model on a large unlabeled collection using a contrasive loss.
     Performance tends to increase when changing the base encoder model to that pretrained with contrastive loss.}
\label{table: MSMARCO.pretrain}
\end{table*}

We conduct experiments to determine if our findings are consistent when using encoder-only base models pretrained with contrastive loss. Specifically, we use the Contriever pretrained model weights~\citep{Izacard2021TowardsUD}. Table~\ref{table: MSMARCO.pretrain} shows that switching to this base model with additional pretraining significantly improves performance, especially in average OOD performance.

Importantly, we see again that our earlier findings persist even with this more powerful base model. We see again that  LoRA shows superior performance in OOD settings but worse performance in in-domain settings. However, we notice that there is LoRA benefits less when trained on Contriever compared to BERT-base. 

When comparing the rate of improvement for average OOD performance and the rate of degradation for in-domain between FT and LoRA, we can see that (1) the rate of improvement in average OOD performance using Contriever (3.36\%) is not as substantial as that using BERT-base (7.24\%) and (2) the improvement rate in OOD (3.36\%) is smaller than the degradation rate than on in-domain settings (3.96\%). 
We hypothesize that this is due to Contriever's adaptation to various domains during its pretraining with a massive unsupervised dataset, making it less likely to over-adapt to a specific training dataset's distribution and maintaining its generalizability.

\section{Do Findings Generalize to Other Training Datasets? A Case Study on Google Natural Questions}

\begin{figure}[t!]
    \centering
    \begin{subfigure}[b]{0.47\textwidth}
        \includegraphics[width=\textwidth]{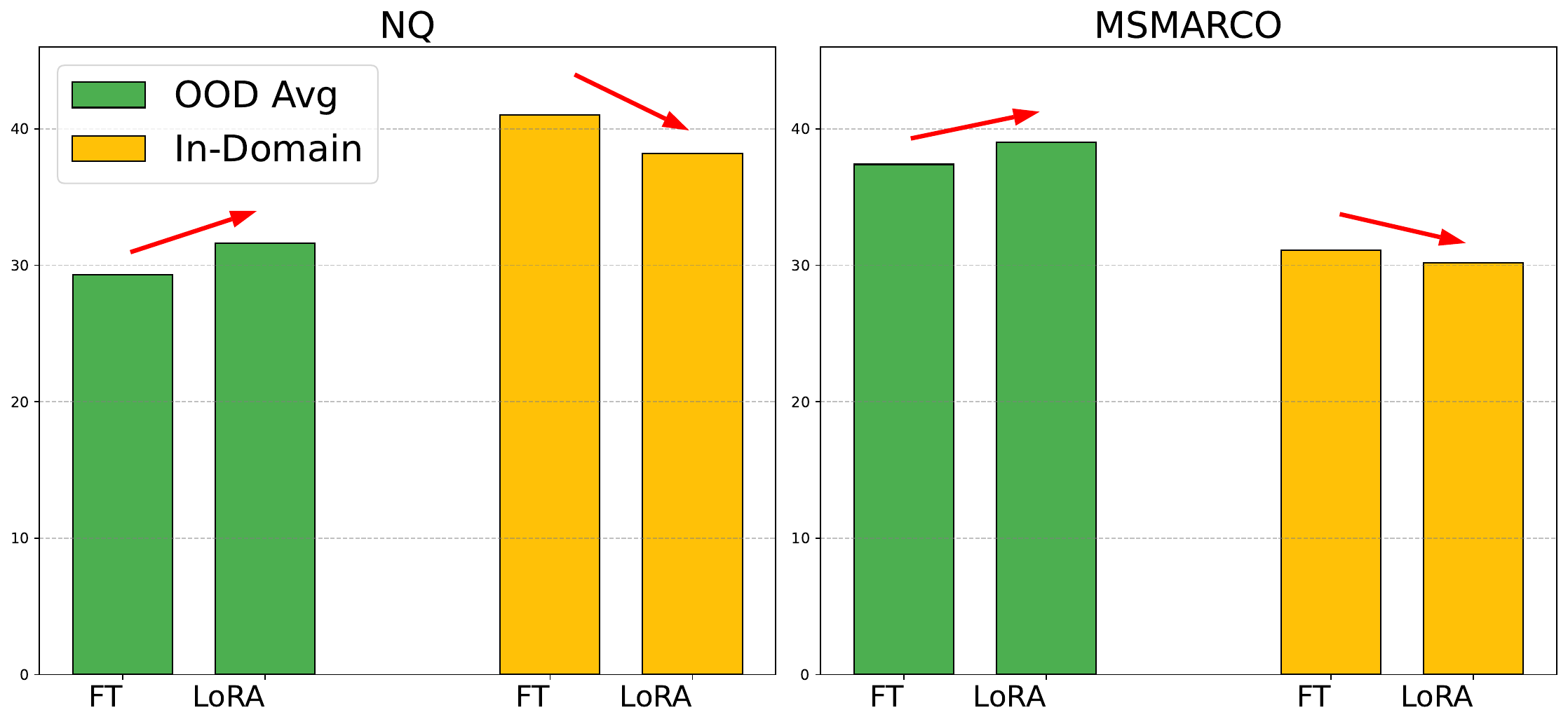}
        \newsubcap{Performance of Asymmetric Dual Encoder when trained with NQ (left). Results tend to be similar to that of MSMARCO (right): LoRA tends to show higher performance in out-of-domain (OOD) but lower performance in in-domain.}
        \label{fig: NQ.peft}
    \end{subfigure}
    \hfill 
    \begin{subfigure}[b]{0.47\textwidth}
        \includegraphics[width=\textwidth]{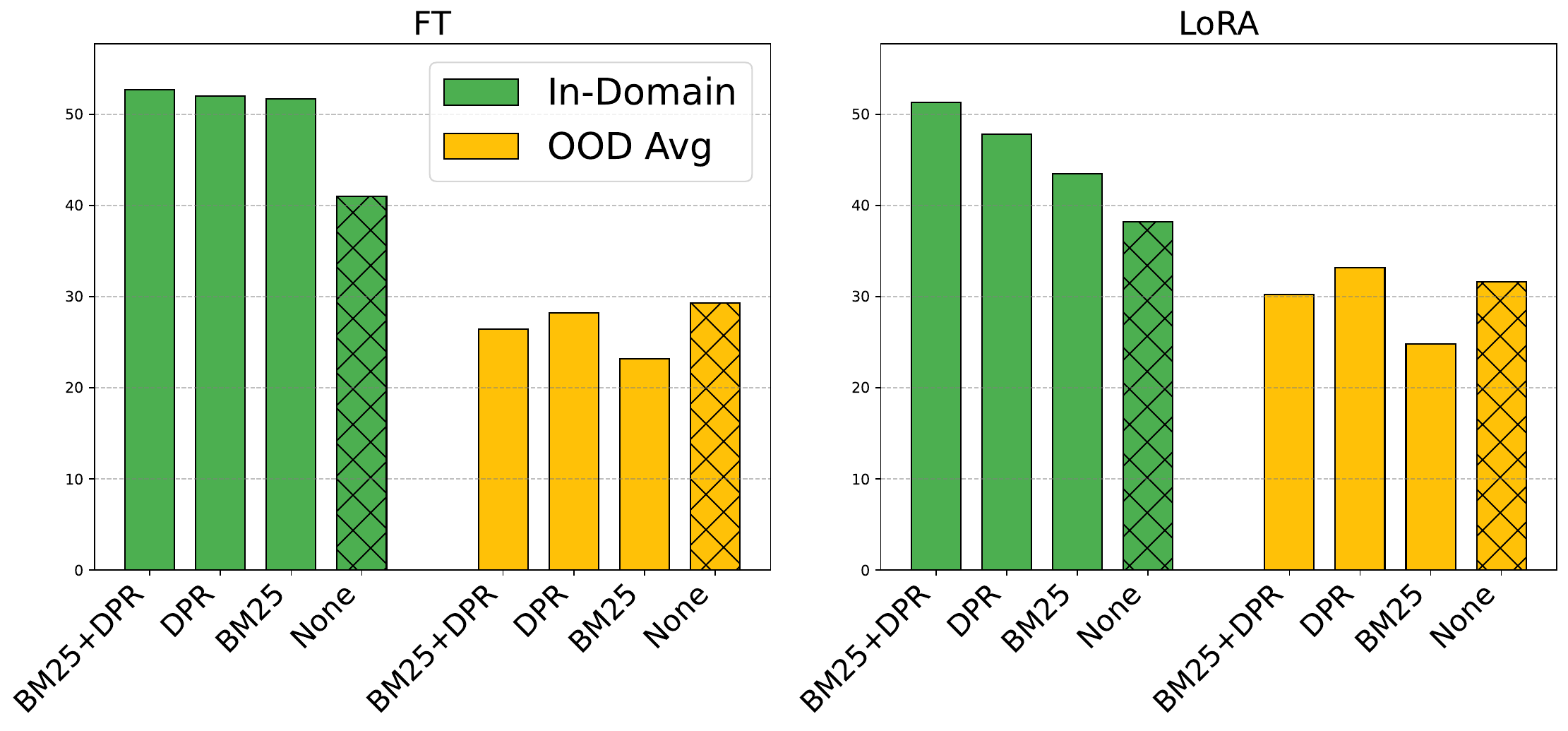}
        \newsubcap{
        Performance of dual encoders trained on the NQ dataset with different negative sampling strategies. 
        Adding hard negatives improves in-domain performance but degrades out-of-domain (OOD) performance for both FT and LoRA.}
        \label{fig: NQ.neg}
    \end{subfigure}
\end{figure}

To analyze whether findings presented in \cref{sec: rq1,sec: rq2,sec: rq3} generalize to different training datasets, we experiment over another widely used retrieval collection, Google Natural Questions~\citep{Kwiatkowski2019NaturalQA} (NQ). 
While NQ is significantly smaller than MSMARCO (NQ is comprised of 307k training examples, while MSMARCO contains over 1M queries and 8.8M passages) and only contains passages extracted from Wikipedia, we note that trends observed on models trained on MSMARCO generally hold constants for NQ, suggesting that they generalize across training datasets. 

\paragraph{Models trained on MSMARCO or NQ have similar in- and out-of-domain performance characteristics}
Figure~\ref{fig: NQ.peft} compares the performance of a dual encoder model trained on MSMARCO and NQ.
Results show similar trends across the datasets: 
full fine-tuning generally improves in-domain performance, but LoRA achieves better nDCG@10 on out-of-domain tasks in BEIR.

\paragraph{Many hard negative mining approaches remain detrimental to out-of-domain performance}
Similarly to MSMARCO in \cref{sec: rq2}, using mined hard negative when training on NQ seems to negatively affect out-of-domain performance. 
Figure~\ref{fig: NQ.neg} compares the effect of negatives on in-domain and out-of-domain performance for FT and LoRA trained asymmetric dual encoders. 
Compared to \cref{sec: rq2}, we experiment with two different rankers to select hard negatives:
BM25 and model-based negatives (DPR\footnote{We use the negatives provided from \url{https://github.com/facebookresearch/DPR}}).
We also evaluate using a combination of both (``BM25+DPR'' in \cref{fig: NQ.neg}).
While all three mining approaches yield improvements in in-domain performance, they are rivaled or bested by using in-batch negative only (``None'' in \cref{fig: NQ.neg}).
This finding is consistent for both FT and LoRA-trained models. 
Overall, our observations support previous research calling for careful selection of hard negatives~\citep{Santhanam2021ColBERTv2EA}.
Further, they highlight how the wrong mining approach is far more likely to hurt out-of-domain performance than in-domain performance, as, in the latter case, incorporating negative examples is consistently beneficial.
 
\begin{figure}[t!]
    \centering
    \includegraphics[width=0.8\textwidth]{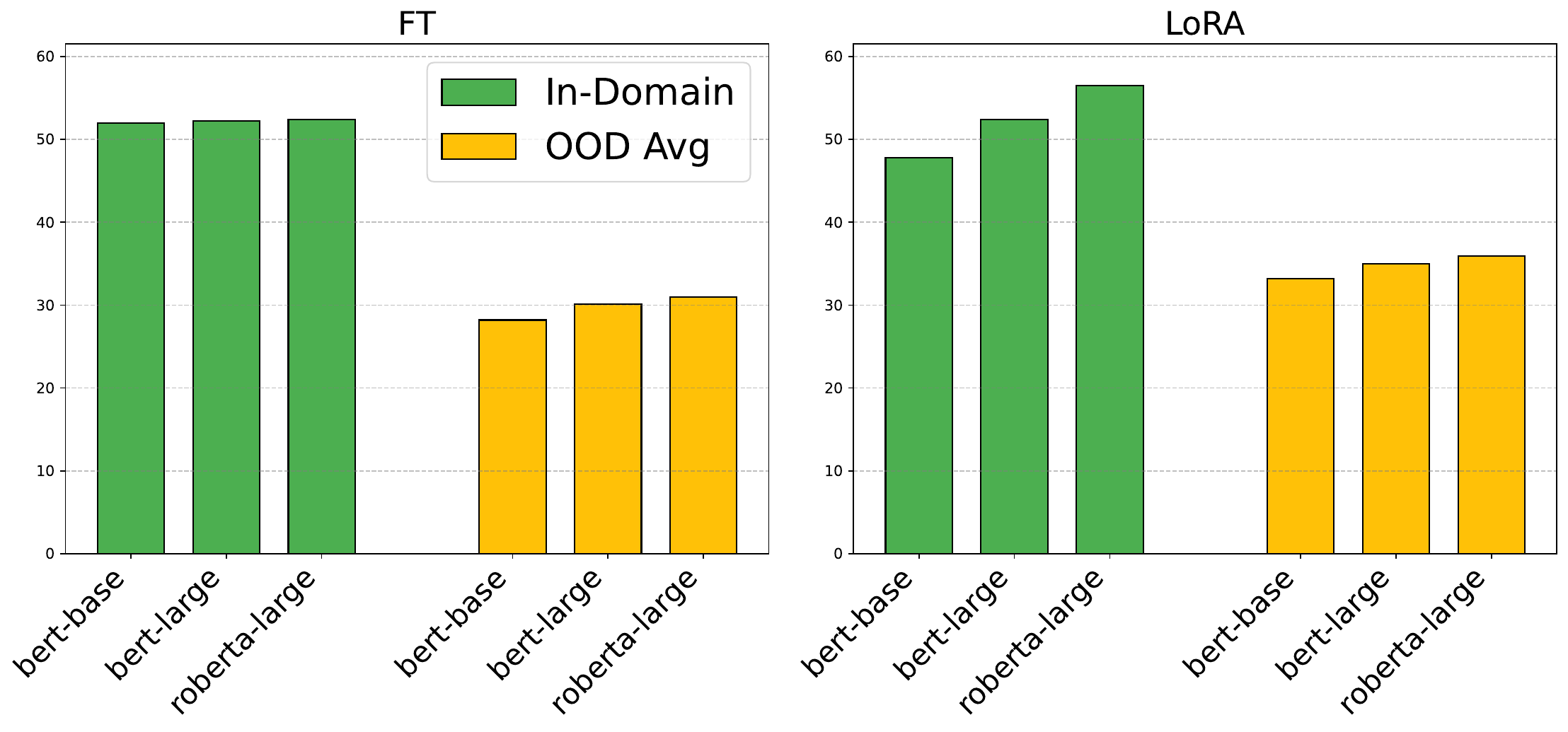}
    \caption{
    Impact of model size on the performance of dual encoders trained on NQ. 
    Larger models consistently lead to higher performance in both in-domain and out-of-domain (OOD). 
    LoRA shows larger gains when moving to larger models.}
    \label{fig: NQ.model_size}
\end{figure}

\paragraph{Advantages of LoRA on out-of-domain generalization is consistent across model sizes.}
Figure~\ref{fig: NQ.model_size} presents the in-domain and out-of-domain performance across various model sizes. We can see that the trend when trained with MSMARCO persists; a superior base model consistently yields enhanced performance in out-of-domain scenarios, with LoRA especially benefiting more from an improved base model. Surprisingly, the data shows that using RoBERTa-large as the base model enables LoRA to exhibit higher performance than fine-tuning (FT), even in the in-domain setting.

\paragraph{MSMARCO vs. NQ as training dataset}
We observed that fine-tuning with MSMARCO consistently yields robust performance (Table~\ref{table: LoRA_vs_full}), surpassing that achieved with NQ (Table~\ref{table: NQ.LoRA_vs_full} in Appendix~\ref{app:nq.qa1}), confirming trends observed by \citet{Thakur2021BEIRAH} and \citet{Ni2021LargeDE}. 
In terms of the average out-of-domain performance reduction rate, models trained with LoRA exhibit a lower reduction rate of 17.4\% compared to when training full parameter (FT), which shows a 20.2\% reduction. To calculate the average out-of-domain performance, we averaged the reduction rate excluding MSMARCO and NQ data. This pattern indicates that LoRA experiences a smaller performance drop with fewer training datasets, a trend consistent with previous studies in parameter efficient methods~\citep{Ustun2022WhenDP, Litschko2022ParameterEfficientNR}.

\section{Conclusion}

In this work, we investigate the impact of training strategies on the generalizability of dense retrieval models. Our focus is on scenarios that avoid extra resource-intensive steps, such as (1) comparing LoRA with full fine-tuning and (2) designing optimal batch sizes for fine-tuning. We further examine how these findings align with other resource-intensive methods for out-of-domain retrieval (\textit{i.e.,} token-level late-interaction models, scaling model size). Across various experiments, we observe a consistent trend: (1) LoRA invariably enhances generalizability, and (2) under identical GPU configurations, increasing the in-batch size typically yields more robust performance compared to adding hard negatives. 
Furthermore, we find that our insights complement popular techniques for boosting out-of-domain performance. Our study offers practical, actionable insights for developing dense retrieval models with high generalizability.

\section{Limitations}

Most of our experiments are conducted on smaller base models relative to some larger choices like Llama~\citep{Ma2023FineTuningLF}. We have demonstrated some robustness of our findings under scaling (\S\ref{sec: rq3}), but further investigation is needed. 

Our study does not explore the compatibility of our approach with data augmentation methods for out-of-domain generalization. Also, we have not investigated whether our approach maintains its effectiveness when trained on a diverse combination of domains, rather than a single training dataset (NQ, MSMARCO).

Our focus is primarily on widely-used negative sampling strategies that do not involve resource-intensive steps like distillation~\citep{Santhanam2021ColBERTv2EA, Formal2021SPLADEVS, Ren2021RocketQAv2AJ}, leading to a lack of exploration in various other negative sampling methods.

\bibliography{iclr2021_conference}
\bibliographystyle{iclr2021_conference}

\appendix
\section{Colbert} \label{app: colbert}
\begin{table*}[ht!]
\centering
\fontsize{6.5}{10}\selectfont
    \begin{tabular}{c|ccccccccccccc|c|c}
    \toprule
    & & \multicolumn{13}{c}{Out-Of-Domain (OOD)} & \multicolumn{1}{c}{In-Domain} \\
    \midrule
    & FI & TR & NF & QU & SD & SF & TO & AR & HO & DB & FE & CL & NQ& \textit{Avg} & MS  \\
\midrule
    Beir & 31.7 & \text{67.7} & 30.5 & \text{85.4} & 14.5 & 67.1 & 20.2 & 23.3 & 59.3 & \text{39.2} & \text{77.1} & 18.4  & \text{52.4} & 45.1& \text{40.1} \\
     Ours & 31.9 & 65.1 & 31.8 & 83.8 & 15.1 & 67.5 & 19.5 & 24.0 & 58.4 & 38.7 & 77.9 & 18.2 & 51.7 & 44.9 & 39.2    \\
    \bottomrule
    \end{tabular}
\caption
     {\fontsize{7.5}{10}\footnotesize ``Ours'' is the performance of Colbert we replicate for fair comparison and ``Beir'' is the performance provided from Table~\ref{table: colbert}, which is widely used. The performance tends to be similar.}
\label{table: colbert 2}
\end{table*}

To study each training configuration's impact and ensure a fair comparison, we replicated the experiment for consistent results (Ours in Table~\ref{table: colbert}). The result aligns closely with the result of Colbert in \citet{thakur2021beir} (Beir in Table~\ref{table: colbert}).

\section{Calculating distribution similarity}
\label{app: domain_similarity}

To explore the impact of hard negatives on datasets within similar distribution, we conduct an analysis using the BEIR dataset, grouping them based on distribution similarity. To assess similarity, our methodology is as follows: First, we sample 50 instances from the corpus of each dataset. Second, we generate embeddings for each instance using contriever~\citep{Izacard2021TowardsUD}. Third, for each instance, we compute its similarity (dot product) with other embeddings, identifying the most relevant dataset, and excluding its original dataset. Fourth, for each dataset, we determine which dataset appears most frequently (out of 50) and regard this as the dataset with the most similar distribution. Last, using this information, we cluster datasets that are interconnected. This process is repeated five times, and we observe that the resulting groupings tend to be consistently similar. 
We categorize datasets grouped with MSMARCO more than three times as having a similar distribution. Consequently, datasets such as trec-covid, NFcorpus, scidocs, scifacts, and arguana are considered to share a similar distribution with MSMARCO. One thing we notice is that datasets with Wikipedia as their source consistently tend to be grouped together, leading us to assume that the grouping shows a high tendency of distribution.

\section{Performance of different model sizes when trained with MSMARCO}
\label{app: MSMARCO.model_size}
Table~\ref{table: MSMARCO.basemodel} presents the overall BEIR performance of various base model sizes of an asymmetric dual encoder trained on MSMARCO, without adding hard negatives. The performance of both in-domain and out-of-domain generally improves with the use of larger encoder models.

\begin{table*}[t!]
\centering
\fontsize{6.5}{10}\selectfont
    \begin{tabular}{c|ccccccccccccc|c|c}
    \toprule
    & \multicolumn{14}{c}{\textbf{Out-Of-Domain (OOD)}}  &  \multicolumn{1}{c}{\textbf{In-Domain}}\\
    \midrule
    & FI & TR & NF & QU & SD & SF & TO & AR & HO & DB & FE & CL & NQ & \emph{Avg} & MS \\
    \midrule
     \multicolumn{14}{l}{\textbf{Bert Small}} \\
    \midrule
    FT & 18.5 & 63.7 & 25.5 & 75.1 & 6.5 & 43.8 & 27.9 & 28.9 & 40.7 & 41.5 & 54.0 & 13.8 & 30.4& 36.2 & 28.9   \\
    LoRA & 18.6 & 71.0 & 25.4 & 74.4 & 7.5 & 46.4 & 28.7 & 29.9 & 42.0 & 42.4 & 56.4 & 14.9 & 28.1 & 37.4 & 27.7  \\
    \midrule
     \multicolumn{14}{l}{\textbf{Bert Medium}} \\
    \midrule
    FT &  18.2 & \underline{72.7} & 25.8 & 74.1 & 6.8 & 45.9 & 27.0 & 28.8 & 41.0 & 41.2 & 54.2 & 12.9 & 31.5& 36.9 & 29.9  \\
    LoRA & 18.8 & \text{71.8} & 26.0 & \underline{77.1} & 7.9 & 47.6 & \textbf{30.0} & 30.6 & 42.1 & 40.7 & 57.1 & 15.3 & 29.8 & 38.1 & 28.6  \\
    \midrule
    \multicolumn{13}{l}{\textbf{Bert Base}} \\
    \midrule
     FT &  18.6 & 64.2 & 25.7 & 72.4 & 7.2 & 45.3 & 26.8 & 31.1 & 44.1 & \underline{44.1} & \underline{57.5} & 13.8 & 35.7& 37.4 & \text{31.1}    \\
     LoRA &  \underline{19.5} & 67.9 & \underline{27.1} & 73.4 & \underline{8.0} & \underline{49.4} & \textbf{30.0} & \underline{33.8}  & \underline{45.4} & 43.5 & \textbf{58.7} & \textbf{15.8} & 34.2 & \underline{39.0} & 30.2\\
    \midrule
    \multicolumn{14}{l}{\textbf{Bert Large}} \\
    \midrule
    FT &  19.3 & 70.0 & 26.5 & 76.7 & 7.5 & 47.6 & \underline{28.1} & 31.5 & 42.5 & 43.5 & 57.0 & 14.1 & \underline{36.1}& 38.5 & \textbf{32.7} \\
    LoRA &  \textbf{23.2} & \textbf{76.7} & \textbf{28.2} & \textbf{79.6} & \textbf{8.5} & \textbf{54.9} & 26.1 & \textbf{34.1} & \textbf{45.9} & \textbf{45.0} & 57.3 & \underline{15.4} & \textbf{36.6} & \textbf{40.9} & \underline{31.4} \\
    \bottomrule
    \end{tabular}
\caption
     {\fontsize{6.5}{10}\footnotesize Overall BEIR performance of different base model sizes of asymmetric dual encoder trained with MSMARCO (without hard negatives). The best and second best over all the model sizes in \textit{bold} and \underline{underline} respectively. }
\label{table: MSMARCO.basemodel}
\end{table*}

\section{Results with NQ as training dataset}
\label{app: NQ}
\subsection{Training Method}
\label{app:nq.qa1}
\begin{table*}[t!]
\centering
\fontsize{6.5}{10}\selectfont
    \begin{tabular}{c|ccccccccccccc|c|c}
    \toprule
     & \multicolumn{14}{c}{\textbf{Out-Of-Domain (OOD)}}  &  \multicolumn{1}{c}{\textbf{In-Domain}}\\
    \midrule
    & FI & TR & NF & QU & SD & SF & TO & AR & HO & DB & FE & CL & MS & \emph{Avg} & NQ \\
    \midrule
        \multicolumn{14}{l}{\textbf{Asymmetric Dual Encoder}} \\
    \midrule
    FT & 16.0 & 61.2 & 20.5 & 31.0 & 7.3 & 41.8 & 24.6 & 21.6  & 34.1 & 40.2 & 46.4 & 15.1& \textbf{21.2} & \text{29.3}  & \textbf{41.0}  \\
    LoRA & \textbf{16.6} & \textbf{63.1} & \textbf{21.2} & \textbf{37.4} & \textbf{8.3} & \textbf{42.6} & \textbf{25.6} & \textbf{22.4}  & \textbf{35.0} & \textbf{41.0} & \textbf{52.4} & \textbf{24.6}& 20.0 & \textbf{31.6} & \text{38.2}   \\  
    \midrule
    \multicolumn{15}{l}{\textbf{Symmetric Dual Encoder}} \\
    \midrule
    FT & 6.1 & \textbf{26.8} & 8.6 & \textbf{75.0} & 4.9 & 34.2 & 2.1 & 3.1  & \textbf{26.5} & 5.7 & 23.1 & \textbf{13.9} & 6.9  & \text{18.2}& \textbf{28.0}  \\
    LoRA & \textbf{7.5} & 22.9 & \textbf{9.8} & 73.7 & \textbf{6.5} & \textbf{36.0} & \textbf{4.9} & \textbf{5.9} & 25.3 & \textbf{7.9} & \textbf{23.4} & 12.8 & \textbf{8.7} & \textbf{18.8}& 25.9   \\
    \bottomrule
    \end{tabular}
\caption
     {\fontsize{7.5}{10}\footnotesize BEIR performance of asymmetric and symmetric dual encoders trained without hard negatives using NQ as the training dataset. Best from FT and LoRA in \textbf{bold}.}
\label{table: NQ.LoRA_vs_full}
\end{table*}

Table~\ref{table: NQ.LoRA_vs_full} shows that parameter-efficient training (PEFT) consistently achieves higher performance in out-of-domain compared to the traditional approach of training full parameters (FT).

\subsection{Batch Design}
\begin{table*}[t!]
\centering
\fontsize{6.5}{10}\selectfont
    \begin{tabular}{cc|ccccccccccccc|c|c}
    \toprule
    && \multicolumn{14}{c}{\textbf{Out-Of-Domain (OOD)}}  &  \multicolumn{1}{c}{\textbf{In-Domain}}\\
    \midrule
    && FI & TR & NF & QU & SD & SF & TO & AR & HO & DB & FE & CL & MS & \emph{Avg} & NQ \\
    \midrule
    \multirow{4}{*}{Full}
    & $R_{BM25+DPR}$ & 13.2 & 42.1 & 17.6 & 28.3 & 5.8 & 32.1 & 24.4 & 20.8  & 33.8 & 40.2 & 48.6 & 16.1 &19.6 & 26.4& \textbf{52.7}   \\
    & $R_{DPR}$ & 11.8 & 54.0 & 19.3 & 30.2 & 6.1 & 35.9 & \textbf{25.6} & \textbf{25.0} & \textbf{34.1} & \textbf{40.7} & 48.6 & 16.4 &19.5& 28.2& 52.0   \\
    & $R_{BM25}$ & 10.4 & 35.5 & 15.8 & 19.2 & 4.9 & 25.9 & 19.1 & 10.8 & 32.9 & 38.0 & \textbf{52.6} & \textbf{18.7} &18.1& 23.2& 51.7    \\
    & $R_{None}$ & \textbf{16.0} & \textbf{61.2} & \textbf{20.5} & \textbf{31.0} & \textbf{7.3} & \textbf{41.8} & 24.6 & 21.6 & \textbf{34.1} & 40.2 & 46.4 & 15.1 &\textbf{21.2} & \textbf{29.3} & 41.0  \\
    \midrule
    \multirow{4}{*}{LoRA}
    & $R_{BM25+DPR}$ & 13.9 & 54.5 & 20.5 & 22.9 & 7.5 & 43.5 & 26.4 & 20.8 & 40.4 & 45.4 & \textbf{55.0} & 21.6 &20.8& 30.2 & \textbf{51.3}   \\
    & $R_{DPR}$ &  15.5 & \textbf{63.9} & 20.7 & \text{37.8} & 7.8 & \textbf{44.5} & \textbf{31.8} & \textbf{25.8} & \textbf{40.8} & \textbf{48.4} & 50.0 & 21.6 &\textbf{22.4}& \textbf{33.2}& 47.8  \\
    & $R_{BM25}$ & 10.2 & 51.0 & 17.2 & 18.5 & 6.3 & 28.2 & 17.8 & 13.9 & 35.8 & 39.1 & 49.1 & 18.1 & 17.1 & 24.8 & 43.5  \\
    & $R_{None}$ & \textbf{16.6} & 63.1 & \textbf{21.2} & \textbf{37.4} & \textbf{8.3} & 42.6 & 25.6 & 22.4 & 35.0 & 41.0 & 52.4 & \textbf{24.6} &20.0 & 31.6 & 38.2   \\
    \bottomrule
    \end{tabular}
\caption
     {\fontsize{6.5}{10}\footnotesize Adding hard negatives consistently enhances in-domain performance (NQ). However, for out-of-domain tasks, adding hard negatives tends to degrade performance unless they are selected very carefully (LoRA $R_{DPR}$). Such results suggest that adding hard negatives makes the model adapt strongly to specific datasets, making it challenging to generalize effectively across different domains. All experiments in the table use the asymmetric dual encoder with NQ as the training dataset.}
\label{table: NQ.negatives}
\end{table*}

\paragraph{Adding Hard Negatives}
Table~\ref{table: NQ.negatives} demonstrates that incorporating hard negatives consistently improves in-domain performance, yet it often reduces out-of-domain performance. Our experiments use negatives from \citep{Karpukhin2020DensePR}, including BM25 negatives ($R_{BM25}$) and model-based negatives ($R_{DPR}$). Please note that the model used for $R_{DPR}$ is DPR~\citep{Karpukhin2020DensePR}, a dense retrieval model known for its superior in-domain performance compared to BM25. $R_{None}$ is when there is no hard negatives used during the training step and only with in-batch negatives. $R_{BM25+DPR}$ is when using a mix of DPR and BM25 negatives hard negatives, effectively doubling the training dataset size\footnote{$R_{BM25+DPR}$ contains twice as many training datasets compared to others since each query includes two instances, one negative from BM25 and one from DPR}. 

In both PEFT and FT experiments, we observe that $R_{BM25+DPR}$ achieves the best in-domain dataset performance, while $R_{None}$ shows the lowest, confirming that hard negatives enhance in-domain dataset performance. However, for out-of-domain datasets, $R_{None}$ performs best, and adding negatives seems to worsen performance. This indicates that hard negatives may cause the model to overfit to a single training dataset distribution, limiting its generalization to out-of-domain datasets.

As observed in various studies on hard negative selection~\citep{Santhanam2021ColBERTv2EA, Formal2021SPLADEVS}, we could also see case where hard negatives mined with high-performance models improve out-of-domain performance (LoRA performance with $R_{DPR}$). Conversely, $R_{BM25}$, which utilizes BM25 for negatives, consistently lowers performance, even when combined with $R_{DPR}$. We speculate that this significant drop with $R_{BM25}$ relates to BEIR's tendency to show high performance with BM25, leading to a higher incidence of false negatives. This finding highlights the critical importance of selecting appropriate hard negatives for out-of-domain performance.

\begin{table*}[t!]
\centering
\fontsize{6.5}{10}\selectfont
    \begin{tabular}{cc|ccccccccccccc|c|c}
    \toprule
    && \multicolumn{14}{c}{\textbf{Out-Of-Domain (OOD)}}  &  \multicolumn{1}{c}{\textbf{In-Domain}}\\
    \midrule
    &batch size& FI & TR & NF & QU & SD & SF & TO & AR & HO & DB & FE & CL & MS & \emph{Avg} & NQ \\
    \midrule
    
    \multirow{2}{*}{Full}
    & B & \text{16.0} & \text{61.2} & \text{20.5} & \text{31.0} & \textbf{7.3} & \text{41.8} & \textbf{24.6} & 21.6 & \text{34.1} & 40.2 & 46.4 & 15.1 &\text{21.2} & \text{29.3} & 41.0  \\
    & $2 \times \text{B}$ & \textbf{17.3} & \textbf{64.6} & \textbf{22.4} & \textbf{33.0} & \text{6.9} & \textbf{45.2} & 23.5 & \textbf{21.9} & \textbf{34.3} & \textbf{41.9} & \textbf{47.1} & \textbf{16.8} & \textbf{21.6} & \textbf{30.5}& \textbf{46.1}  \\  
    \midrule
    \multirow{3}{*}{LoRA}
    & B & \text{16.6} & 63.1 & \text{21.2} & 37.4 & \textbf{8.3} & 42.6 & 25.6 & 22.4 & 35.0 & 41.0 & 52.4 & \text{24.6} &20.0 & 31.6 & 38.2   \\
    & $2 \times \text{B}$ & 17.0 & 63.9 & 21.7 & 39.7 & 7.7 & 44.6 & \textbf{32.1} & \textbf{23.1} & 35.7 & \textbf{46.7} & \textbf{52.6} & 25.4 & 21.3 & 33.2 & 41.6  \\
    & $4 \times \text{B}$ & \textbf{17.6} & \textbf{64.8} & \textbf{22.5} & \textbf{42.8} & \textbf{8.3} & \textbf{45.8} & 29.9 & \textbf{23.1} & \textbf{35.8} & 43.9 & 51.8 & \textbf{26.0} & \textbf{22.4}& \textbf{33.4} & \textbf{42.4}   \\ 
    \bottomrule
    \end{tabular}
\caption
     {\fontsize{6.5}{10}\footnotesize Increasing batch size consistently improves both in-domain and out-of-domain performance. Experiments are conducted with asymmetric dual encoders, without hard negatives, and using NQ as a training dataset.}
\label{table: NQ.inbatch}
\end{table*}

\paragraph{Increasing Batch Size}
Table~\ref{table: NQ.inbatch} shows that increasing batch size improves both in-domain and out-of-domain performance. Such results suggest that when given the same GPU configuration, increasing batch size would be a good option to further improve performance rather than adding hard negatives.

\subsection{Robustness of results across different resource-intensive methods}
\begin{table*}[t!]
\centering
\fontsize{6.5}{10}\selectfont
    \begin{tabular}{cc|ccccccccccccc|c|c}
    \toprule
    && \multicolumn{14}{c}{\textbf{Out-Of-Domain (OOD)}}  &  \multicolumn{1}{c}{\textbf{In-Domain}}\\
    \midrule
    && FI & TR & NF & QU & SD & SF & TO & AR & HO & DB & FE & CL & MS & \emph{Avg} & NQ \\
    \midrule
    
    \multicolumn{14}{l}{\textbf{Bert Base}} \\
    \midrule
    \multirow{2}{*}{Full}
    & $R_{DPR}$ & 11.8 & 54.0 & 19.3 & 30.2 & 6.1 & 35.9 & 25.6 & 25.0 & 34.1 & 40.7 & 48.6 & 16.4 &19.5& 28.2 & \textbf{52.0}   \\
    & $R_{None}$ & 16.0 & 61.2 & 20.5 & 31.0 & 7.3 & 41.8 & 24.6 & 21.6 & 34.1 & 40.2 & 46.4 & 15.1 &21.2 & 29.3 & 41.0  \\
    \midrule
    \multirow{2}{*}{LoRA}
    & $R_{DPR}$ &  15.5 & \textbf{63.9} & 20.7 & \textbf{37.8} & 7.8 & \textbf{44.5} & \textbf{31.8} & \textbf{25.8} & \textbf{40.8} & \textbf{48.4} & 50.0 & 21.6 & \textbf{22.4} & \textbf{33.2} & 47.8  \\
    & $R_{None}$ & \textbf{16.6} & 63.1 & \textbf{21.2} & 37.4 & \textbf{8.3} & 42.6 & 25.6 & 22.4 & 35.0 & 41.0 & \textbf{52.4} & \textbf{24.6} & 20.0 & 31.6 & 38.2   \\ 
    \midrule
    \multicolumn{14}{l}{\textbf{Bert Large}} \\
    \midrule
    \multirow{2}{*}{Full}
    & $R_{DPR}$ & 14.9 & 46.6 & 20.1 & 59.0 & 6.1 & 40.5 & 20.9 & 21.3 & 34.6 & 37.8 & 46.1 & 15.2 &19.8& 30.1 & 52.2  \\
    & $R_{None}$ & 14.8 & 41.1 & 22.2 & \text{57.6} & 7.7 & 44.3 & 24.3 & 31.1 & 34.9 & 38.9 & 44.7 & 16.9 &21.5 & 30.8 & 43.6 \\
    \midrule
    \multirow{2}{*}{LoRA}   
    & $R_{DPR}$ & \textbf{18.2} & \textbf{58.9} & \textbf{23.3} & \textbf{60.5} & \textbf{7.8} & \textbf{48.2} & \textbf{28.2} & \textbf{31.6}& \textbf{42.2} & \textbf{40.2} & \textbf{51.4} & 22.2 & \textbf{22.9}  & \textbf{35.0} & \textbf{52.4} \\
    & $R_{None}$ & 18.1 & 58.4 & 22.3 & 58.9 & 7.5 & 46.9 & 26.9 & 31.3 & 35.7 & 34.7 & 50.0 & \textbf{24.9} & 20.6  & 33.6 & 42.8  \\
    \midrule
    \multicolumn{14}{l}{\textbf{RoBERTa Large}} \\
    \midrule
    \multirow{2}{*}{Full}
    & $R_{DPR}$ & 18.1 & 40.7 & 22.0 & 70.2 & 6.4 & 39.4 & 23.5 & 35.1 & 30.4 & 35.2 & 42.8 & 18.6 & 20.4  & 31.0 & 52.4 \\  
    & $R_{None}$ & 19.5 & 41.4 & 24.4 & 72.6 & \textbf{7.4} & 46.2 & 22.0 & 32.3 & 32.2 & 35.3 & 40.8 & 17.9 & 22.1 & 31.9 & 44.4 \\
    \midrule
    \multirow{2}{*}{LoRA}
    & $R_{DPR}$ & \textbf{21.8} & \textbf{51.8} & \textbf{25.4} & 71.8 & 7.2 & \textbf{48.5} & \textbf{24.8} & \textbf{37.8} & \textbf{38.8} & \textbf{41.2} & \textbf{48.4} & 25.6 & \textbf{23.2}  & \textbf{35.9} & \textbf{56.5}\\  
    & $R_{None}$ & 21.6 & 51.7 & 24.3 & \textbf{73.4} & 7.2 & 43.4 & 23.9 & 37.2 & 35.7 & 40.7 & 46.6 & \textbf{25.7} & 21.8 & 34.9  & 46.1 \\
    \bottomrule
    \end{tabular}
\caption
     {\fontsize{6.5}{10}\footnotesize Performance of different sizes of asymmetric dual encoder trained with NQ without hard negatives.}
\label{table: NQ.basemodel}
\end{table*}

\paragraph{(1) Increasing model size}
Table~\ref{table: NQ.basemodel} shows that both in-domain and out-of-domain performance tends to increase with model size and better base model. Having a better base model consistently leads to higher performance in out-of-domain. LoRA tends to gain more benefits from a better base model, demonstrating higher improvements when considering the average over total scores and even outperforming in-domain scenarios.

\begin{table*}[t!]
\centering
\fontsize{6.5}{10}\selectfont
    \begin{tabular}{c|ccccccccccccc|c|c}
    \toprule
     & \multicolumn{14}{c}{\textbf{Out-Of-Domain (OOD)}}  &  \multicolumn{1}{c}{\textbf{In-Domain}}\\
    \midrule
    & FI & TR & NF & QU & SD & SF & TO & AR & HO & DB & FE & CL & MS & \emph{Avg} & NQ \\
    \midrule
    FT & 13.5 & \textbf{30.4} & 23.0 & \textbf{72.8} & 9.2 & 64.6 & 4.4 & \textbf{37.9} & \textbf{42.0} & 14.1 & 38.8 & \textbf{11.3} & 10.9& 28.7  & \textbf{25.7}   \\ 
    LoRA & \textbf{16.9} & 28.3 & \textbf{25.3} & \textbf{72.8} & \textbf{9.3} & \textbf{64.8} & \textbf{5.4} & 35.2 & 40.1 & \textbf{15.7} & \textbf{40.3} & 11.7 & \textbf{12.8} & \textbf{29.1} & 23.7  \\ 
    \bottomrule
    \end{tabular}
\caption
     {\fontsize{7.5}{10}\footnotesize Performance of Colbert trained with hard negatives sampled from DPR model and using NQ as the training dataset.}
\label{table: NQ.Colbert}
\end{table*}

\paragraph{(2) Use of token-level late-interaction models}
Table~\ref{table: NQ.Colbert} shows the result of Colbert~\citep{Khattab2020ColBERTEA}, a widely used token-level late-interaction model, performance when trained with NQ. The trend tends to persist; LoRA shows higher performance on out-of-domain whereas lower performance in the in-domain dataset.

\end{document}